\documentclass{article}

 \PassOptionsToPackage{numbers, compress}{natbib}
\usepackage[final]{nips_2017}

\usepackage[utf8]{inputenc} 
\usepackage[T1]{fontenc}    
\usepackage{hyperref}       
\usepackage{url}            
\usepackage{booktabs}       
\usepackage{amsfonts}       
\usepackage{nicefrac}       
\usepackage{microtype}      
\usepackage{graphicx}
\usepackage[caption=false,font=footnotesize]{subfig}
\usepackage{tikz}
\graphicspath{ {figures/} }

\title{A visual search engine for Bangladeshi laws}

\author{
  Manash Kumar Mandal \\
  Department of EEE \\
  Khulna University of Engineering \& Technology \\
  Khulna, Bangladesh \\
  \texttt{manashmndl@gmail.com} \\
  \And 
  Pinku Deb Nath \\
  Department of ECE\\
  North South University\\
  Dhaka, Bangladesh \\
  \texttt{pinku.nath@northsouth.edu} \\
  \And 
  Arpeeta Shams Mizan \\
  Department of Law \\
  Dhaka University \\
  Dhaka, Bangladesh \\
  \texttt{amizan@llm15.law.harvard.edu}
  \And 
  Nazmus Saquib \\
  Media Lab \\
  Massachusetts Institute of Technology \\
  Cambridge, MA, USA \\
  \texttt{saquib@mit.edu}
}

\begin{document}

\maketitle

\begin{abstract}

Browsing and finding relevant information for Bangladeshi laws is a challenge faced by all law students and researchers in Bangladesh, and by citizens who want to learn about any legal procedure. Some law archives in Bangladesh are digitized, but lack proper tools to organize the data meaningfully. We present a text visualization tool that utilizes machine learning techniques to make the searching of laws quicker and easier. Using Doc2Vec to layout law article nodes, link mining techniques to visualize relevant citation networks, and named entity recognition to quickly find relevant sections in long law articles, our tool provides a faster and better search experience to the users. Qualitative feedback from law researchers, students, and government officials show promise for visually intuitive search tools in the context of governmental, legal, and constitutional data in developing countries, where digitized data does not necessarily pave the way towards an easy access to information.
\end{abstract}

\section{Introduction}
The digitization of all governmental constitutions and legal documents in Bangladesh started in 2010 as part of a digitization effort from the government. The aim of the digitization program was to make information accessible to every citizen in the country through websites and other forms of online access. With cheaper Internet options and availability of Internet access devices, the vision was to digitize governmental services by 2021 \citep{digitalbd}. However, by interviewing law researchers and students who use the government laws website \citep{bdlaws} on a regular basis, we found that searching for relevant information is quite hard even for veteran researchers (section \ref{sec:user_interviews}). In this paper, we present a visual search tool that utilizes machine learning techniques to provide an intuitive interface to law researchers and amateur students. We received promising feedback on this additional layer of intuitive information access during our pilot studies. 

In section \ref{sec:system_design}, we briefly describe our initial interview process and participants, and the system design process that utilizes existing machine learning tools. Section \ref{sec:data} describes the process of extracting, mining, and visualizing the data. The visual interface and qualitative user feedback are described in section \ref{sec:vis_interface}. We conclude by discussing the potential of using machine learning based intuitive visualization tools as an effective way to augment digital data in developing countries, and the lack of region specific machine learning tools that need to be developed first to create a better experience for native Bangladeshi users.

\section{System design}\label{sec:system_design}

Our system was designed over one year, iterating on our visual and machine learning components based on periodic user feedback. We will briefly summarize the user interview and design iteration process in this section.

\subsection{User interviews}\label{sec:user_interviews}

The initial interview participants were three (academic) law researchers and five law undergraduate students at different stages of their degrees. The interviews were aimed towards understanding the problems associated with using the available text based website in accomplishing research or homework tasks. There were a few recurring themes present across all participant feedback. Most participants also noted that citizens who are dealing with legal procedures in their day to day lives also tend to be consumers of such digitized information.

Finding associated law articles when exploring a single article is where users typically spend most of their time. Usually, there are many hyperlinks in a given article that cite other law articles. Why and where such citations are mentioned are common confusions. Exploring the cited laws in detail is also a tedious process and involves opening and navigating many tabs.

Students who are new to legal terminology -- and also occasionally researchers -- want to understand what are the main entities involved with a particular law. This boiled down to three 'Wh' questions during the interviews: ‘Who’, ‘What’, and ‘When’. For more commonly used laws, researchers tend to remember the relevant entities and amendment timeline better than students. However, for many of the lesser used laws, these are the questions a reader would want to answer quickly to determine if the particular law article is useful. Additionally, a summarization scheme would be useful to glance through the text.

\subsection{Design decisions}

Over a few design iterations and feedback sessions with the initial interview participants, we designed a system that takes account of the themes mentioned by the users. We decided to create a text search system based on any keyword. This takes account of users who are not familiar with legal terminology, and can type in any relevant keyword of their choice. Next, we visualize all relevant search results (law articles) in an interactive citation network diagram, and provide the user the ability to explore each node separately to explore associated entities and themes of the law.

\section{Data mining and visualization techniques}\label{sec:data}

We faced a number of challenges when scraping and mining the corpus to create ngram models. The HTML tag structures were inconsistent, and required us to write custom regular expressions for many special cases. Amendment histories had to be scraped from footnotes of each article separately. Currently, only constitutional laws are available online, and other laws are being digitized by the government's information access office. Our tool is based on the 1100 law articles available in the government website. 

\subsection{Training language models}

Searching based on keywords required us to train bigram and trigram models on the corpus. These models are used when a user chooses to search based on keyword based phrases instead of specific legal terms or act codes.

\subsection{Citation network layout}

Based on a search term, we index the relevant laws and look up our hyperlink database to create a citation network. To create an intuitive layout where similar laws based on topical similarity are placed near one another, we use a graph embedding technique. Document vectors were prepared from scrapped law articles using the doc2vec \citep{le2014distributed} paragraph embedding algorithm. Then we applied manifold learning (multidimensional scaling) to embed the 300 dimensional document vectors onto a 2D vector space. The nodes were placed according to the 2D vector coordinates.

\subsection{Answering 'Wh' questions}
We tackle the 'Wh' questions by using a named entity recognizer (Stanford NER) to extract organization/person names, dates, and locations in each article. Additionally, we include a word cloud of important keywords based on frequency in the language models. Selecting a word in the cloud dynamically updates the article sections in a highlighting mode showing the locations of the selected word. The recent law articles are in the native language (Bangla), and reliable tools for Bangla named entity recognition do not yet exist. This constrained our tool capabilities, and currently around 800 English laws can be explored for entities.

\section{Visual Interface}\label{sec:vis_interface}

\begin{figure}[h]

        \centering
        \subfloat[     ]		{%
      		\includegraphics*[width=0.38\textwidth]{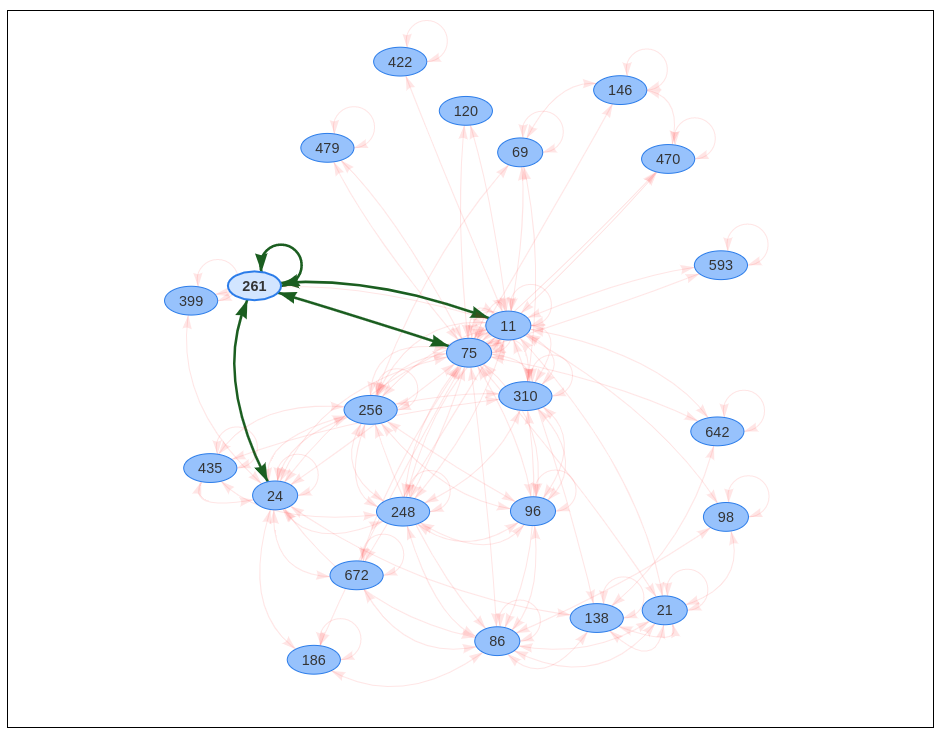}
      		\label{fig:close}
    	}
        \subfloat[]		{%
      		\includegraphics*[width=0.62\textwidth]{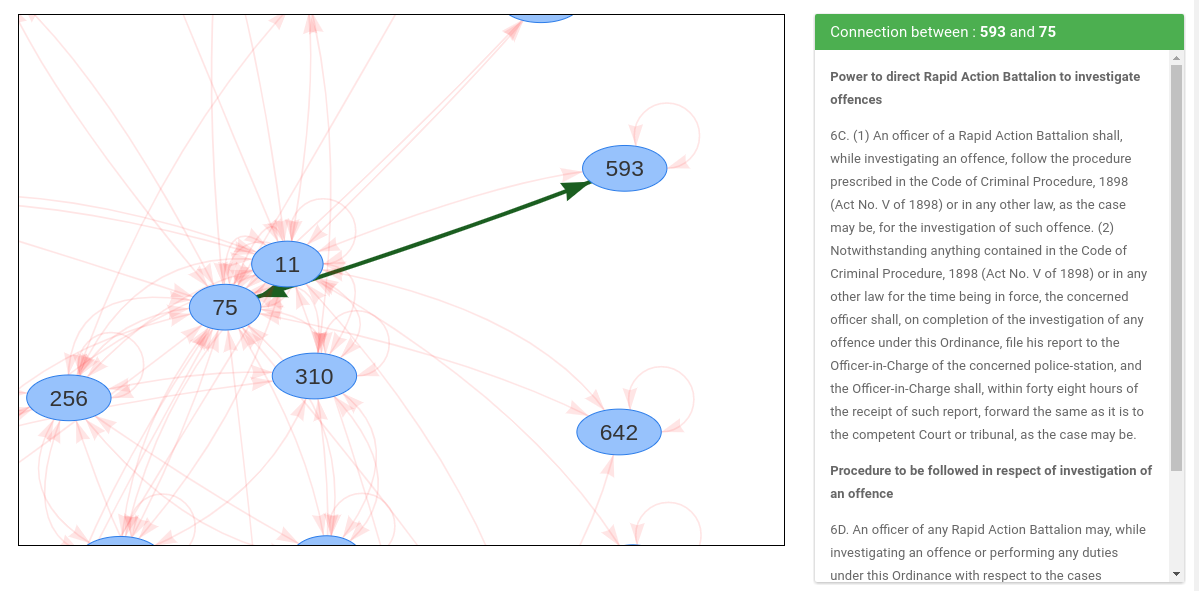}
      		\label{fig:conn}
    	}
    \caption{Visual clues for exploring a citation network. (a) Laying out search results based on doc2vec, (b) edge selection capability to explore the text portion where a citation occurs.}
    \label{fig:closeup}
\end{figure}

Figure \ref{fig:closeup} shows the interactive visual clues used to assist the user in exploring law relationships. Doc2vec based layout algorithm helps the user understand topical similarity between the articles, and interactive selection of edges shows highlighted texts where the citation occurred.

\begin{figure}[h]
   
        \centering
        \includegraphics[width=.7\textwidth]{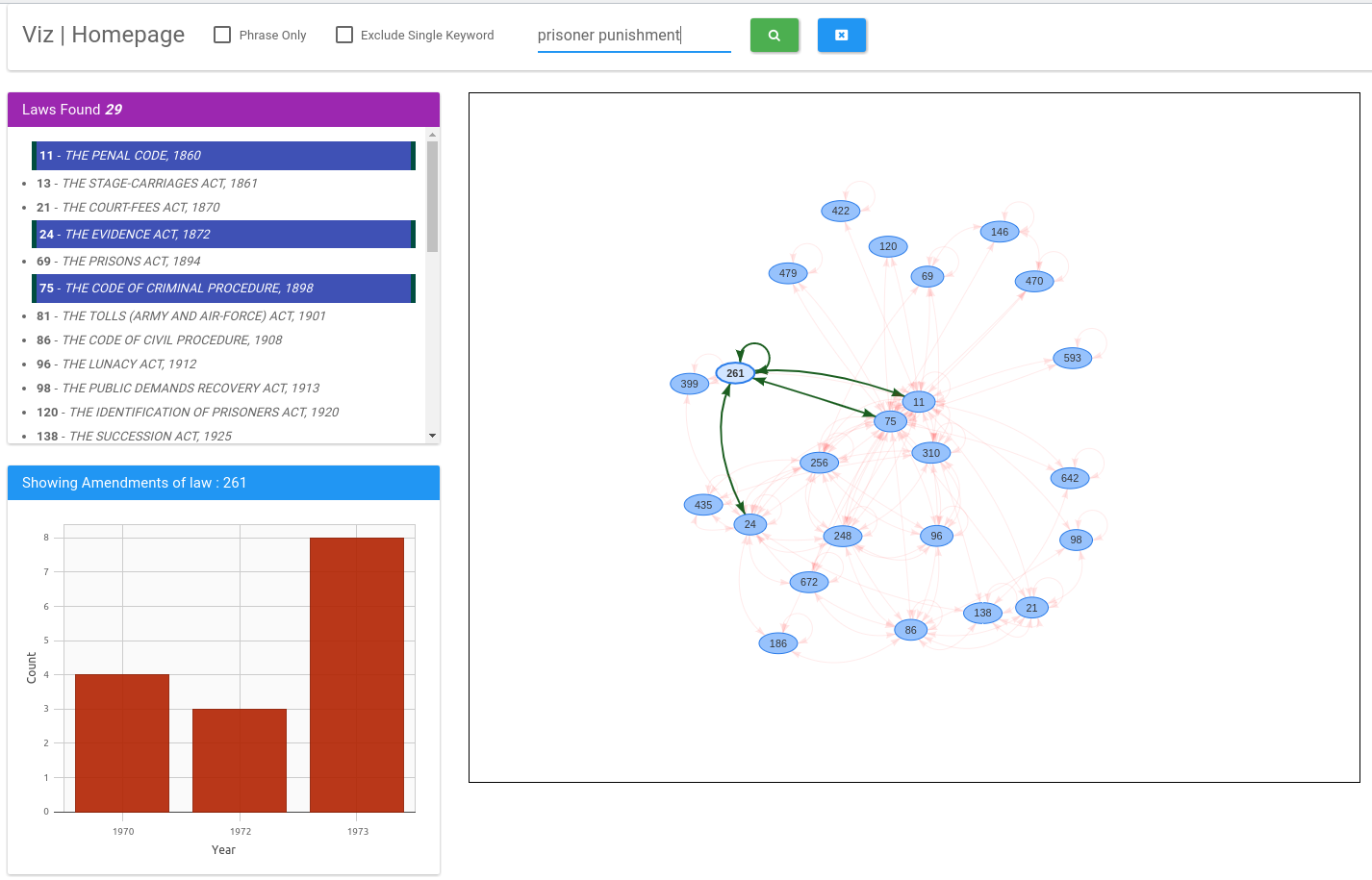}
    
    \caption{Search engine interface.}
    \label{fig:searchview}
\end{figure}

Figure \ref{fig:searchview} shows the search interface where multiple nodes can be selected and explored together (both by clicking the nodes or selecting from the list), and additionally we show a timeline of amendments for the selected nodes. Each node can be double clicked to explore the entity summary view.

\begin{figure}[h]
   
        \centering
        \includegraphics[width=.82\textwidth, height = 0.5\textwidth]{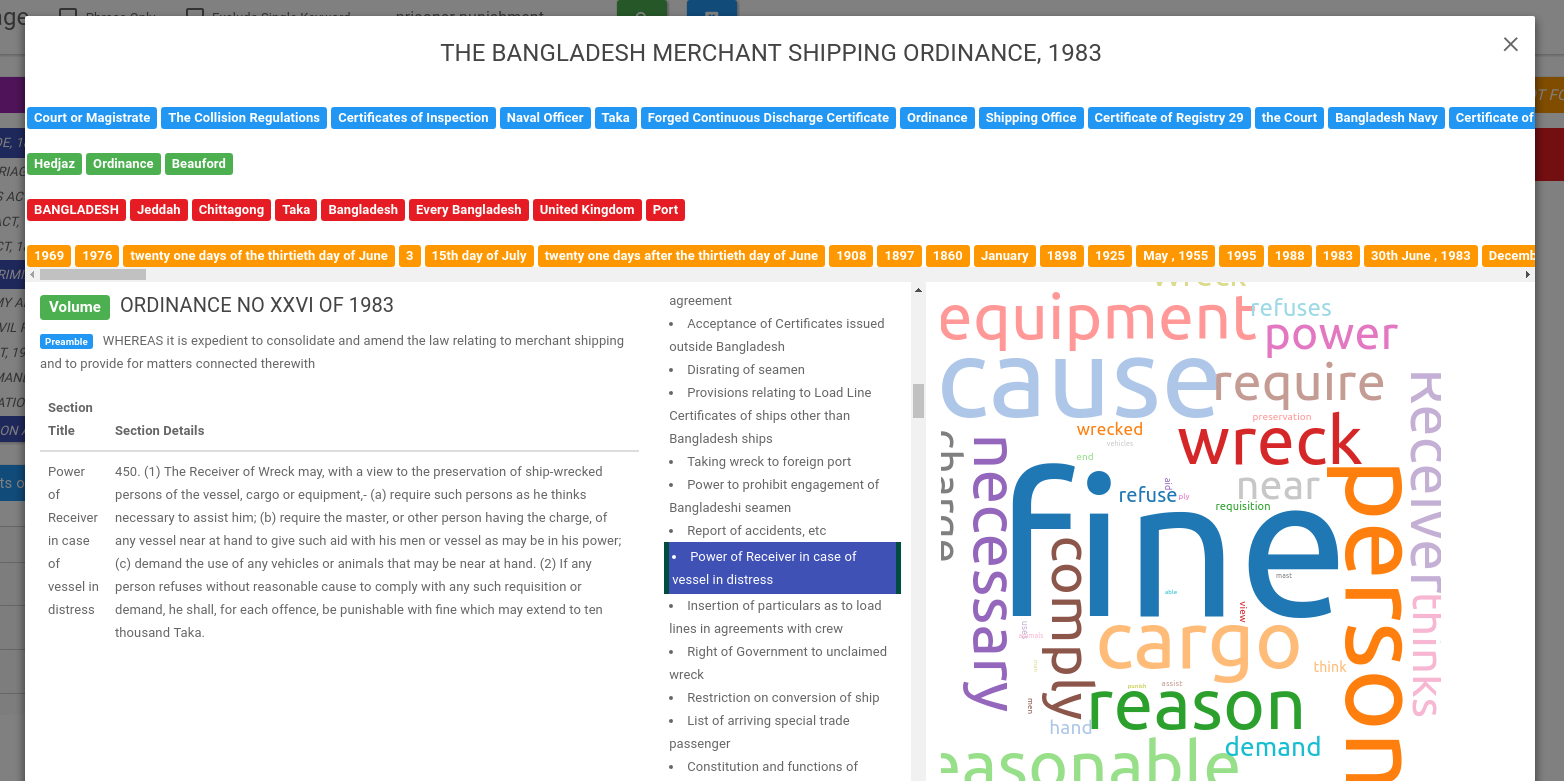}
    
    \caption{A law article's named entities exploration.}
    \label{fig:lawview}
\end{figure}

Figure \ref{fig:lawview} shows the entity summary view. Organizations, locations, and any time related information in the article are shown as clickable tabs. Clicking a tab shows the sections where the tab entity occurs. Sections can be individually selected from the section list, and a word cloud allows the user to highlight section texts where these keywords occur.

We ran a pilot study with a group of 15 lawyers and undergraduate students to evaluate the efficacy of the tool. The feedback we received was very positive. The citation network, named entities, and word clouds as visual signatures to explore laws were well appreciated. However, the inability to explore entities in Bangla laws was the main drawback that some participants pointed out.

\section{Conclusion}

Our research is a work in progress, and the main challenge lies in the unavailability of digitized laws and a good named entity recognition tool for Bangla. We are planning to extend our datasets by including information about Bangladeshi laws that are only available as written texts in books. This will involve collaboration with a number of law departments and government institutions in Bangladesh. A separate research project is currently being led by our team to build a comprehensive NER tagger for Bangla. With more data and good classifiers, we aim to make the visual search engine experience better.

In our interactions with different legal and government institutes during the development phase of the tool, we have seen a lot of enthusiasm among people towards the idea of using a combination of visualization and machine learning techniques to make textual information more accessible to non-experts. Bangladeshi government has the capability to only employ a limited number of experts who can offer proper information and assistance in a wide range of tasks, such as property registration or refugee registration. Some officials remarked that mistakes and confusions frequently arise due to citizen's lack of access to legal information and is one of the main bottlenecks slowing down governmental processes. Our work is a step towards visualizing legal information, and requires more user interviews and design iterations to make it accessible to general citizens. However, our pilot studies and interviews suggest that such directions are worth exploring. With proper user training, visual tools have the potential to bridge the gap towards true advantages of information access.

\bibliographystyle{agsm}
\bibliography{ref}

\end{document}